\documentclass[utf8]{FrontiersinHarvard_nologo} 

\usepackage{url,lineno,microtype,subcaption,lscape}
\usepackage[onehalfspacing]{setspace}
\usepackage{booktabs}
\usepackage{mciteplus}
\usepackage{pdflscape}
\usepackage{rotating}
\usepackage{float}
\usepackage{adjustbox}

\def\keyFont{\fontsize{8}{11}\helveticabold }
\def\firstAuthorLast{Rengel {et~al. }} 
\def\Authors{Rengel, M.\,$^{1,*}$,  and Krumstroh, E.\,$^{1,2}$ }
\def\Address{$^{1}$Max-Planck-Institut für Sonnensystemforschung,  Justus-von-Liebig-Weg 3, 37077, Göttingen, Germany \\
    $^{2}$Georg-August-Universit\"at G\"ottingen,  Faculty of Geoscience and Geography, Geoscience Center, Goldschmidtstr. 3, 37077 Göttingen, Germany}
\def\corrAuthor{Rengel, M.}
\def\corrEmail{rengel@mps.mpg.de}

\begin{document}
\onecolumn
\firstpage{1}
\title{Revisiting a \textbf{historically} suspected impact structure in the Venezuelan Guiana Shield using SRTM topography}
\author[\firstAuthorLast ]{\Authors} 
\address{} 
\correspondance{} 
\extraAuth{}
\maketitle
\begin{abstract}
\tiny
During the 1980s, aerial observations over the Venezuelan Guiana Shield led to the hypothesis that a circular geomorphological feature might represent an impact structure. Radar imagery was subsequently requested and circulated among researchers, but the interpretation remained unresolved. 
Recently rediscovered correspondence and imagery enabled a re-evaluation using modern digital elevation models from the Shuttle Radar Topography Mission (SRTM). The higher-resolution topographic data indicate that the structure lacks diagnostic morphologies of impact craters and is consistent with the Nuria ring dike intrusive complex. Comparison with analogous circular intrusive complexes, including the Kondyor massif, illustrates how circular morphology alone is insufficient for impact identification. 
This study documents a historical episode of geomorphological interpretation prior to the availability of modern digital elevation datasets and illustrates how improved topographic information and geological context can clarify the origin of ambiguous circular landforms relevant to planetary surface analysis.
 \keyFont{ \section{Keywords:} impact craters, remote sensing, geomorphology, Guiana Shield, SRTM, ring dike, planetary analogs}
\end{abstract}
\section{Introduction}

Circular landforms have long attracted attention in planetary science because impact craters represent primary chronological markers on solid planetary surfaces. Consequently, terrestrial circular structures have frequently been investigated as potential impact sites \textcolor{black}{\citep{di2006non,mccall2009half,holmqvist2020lycksele,macgregorsecond,https://doi.org/10.1111/maps.14103}}.

During the 1980s, a circular feature observed during a flyover in the Venezuelan Guiana Shield prompted informal consideration among researchers that it might represent a meteorite impact structure. Radar imagery was requested and circulated among collaborators; however, the available radar data primarily emphasized planform geometry and lacked quantitative topographic information. Consequently, key morphological criteria such as continuous raised rims, bowl-shaped depressions, central uplifts, and ridge continuity could not be reliably evaluated, limiting discrimination between impact and intrusive geomorphological structures \citep{Kenkmann2021}.

The rediscovery of archival correspondence and radar imagery from the original 1980s investigation allows re-evaluation of this early remote-sensing interpretation using modern topographic datasets.
The purpose of this study is to document a historical episode of geomorphological interpretation based on archival radar imagery and to reassess the feature using modern digital elevation models and published geological information, this is a relevant topic to planetary surface interpretation. 

The present study originated from rediscovered correspondence accompanying a radar image circulated among researchers in the early 1980s (L. Maupomé, private communication to G. Bruzual, circa 1981--1983) and does not represent a formal scientific investigation or publication effort. According to the accompanying note preserved by G. Bruzual, the feature was first recognized during a flyover as an unusual geomorphological formation within densely forested terrain. Radar imagery—reportedly originating from NASA sources—was subsequently requested to assess whether the circular landform might indicate a meteorite impact crater. At the time, the apparent circularity and isolated geomorphological expression within densely vegetated terrain motivated comparison with known impact structures visible in radar imagery. However, the limited resolution and lack of quantitative topographic information prevented reliable distinction between impact-related and non-impact circular geomorphological structures at the time.

The historical image corresponds to an airborne real-aperture side-looking radar (SLAR-type) mapping mosaic consistent with regional radar surveys conducted over northern South America during the late twentieth century. 

These real-aperture radar systems produced oblique backscatter imagery with spatial resolutions of several tens of meters and were widely used for geological mapping in tropical regions because they penetrate cloud cover and vegetation canopy.
The original archival radar mosaic could not be georectified because metadata regarding acquisition geometry and projection parameters were unavailable. Consequently, the historical imagery is used qualitatively to document the appearance and interpretation context of the feature rather than for quantitative spatial comparison.
However, the absence of true topographic information and the presence of illumination effects such as foreshortening and brightness contrasts enhanced the apparent circularity of the landform and likely contributed to its initial impact interpretation.

The annotated mosaic identified two anomalous structures considered potentially related to impact processes, motivating re-evaluation using modern datasets. The primary marked feature was interpreted as a possible meteorite crater, while a secondary nearby structure was also noted.

\section{Historical Material and Methods}
\subsection{Archival material}
\begin{figure}
    \centering
    \includegraphics[width=0.6\linewidth]{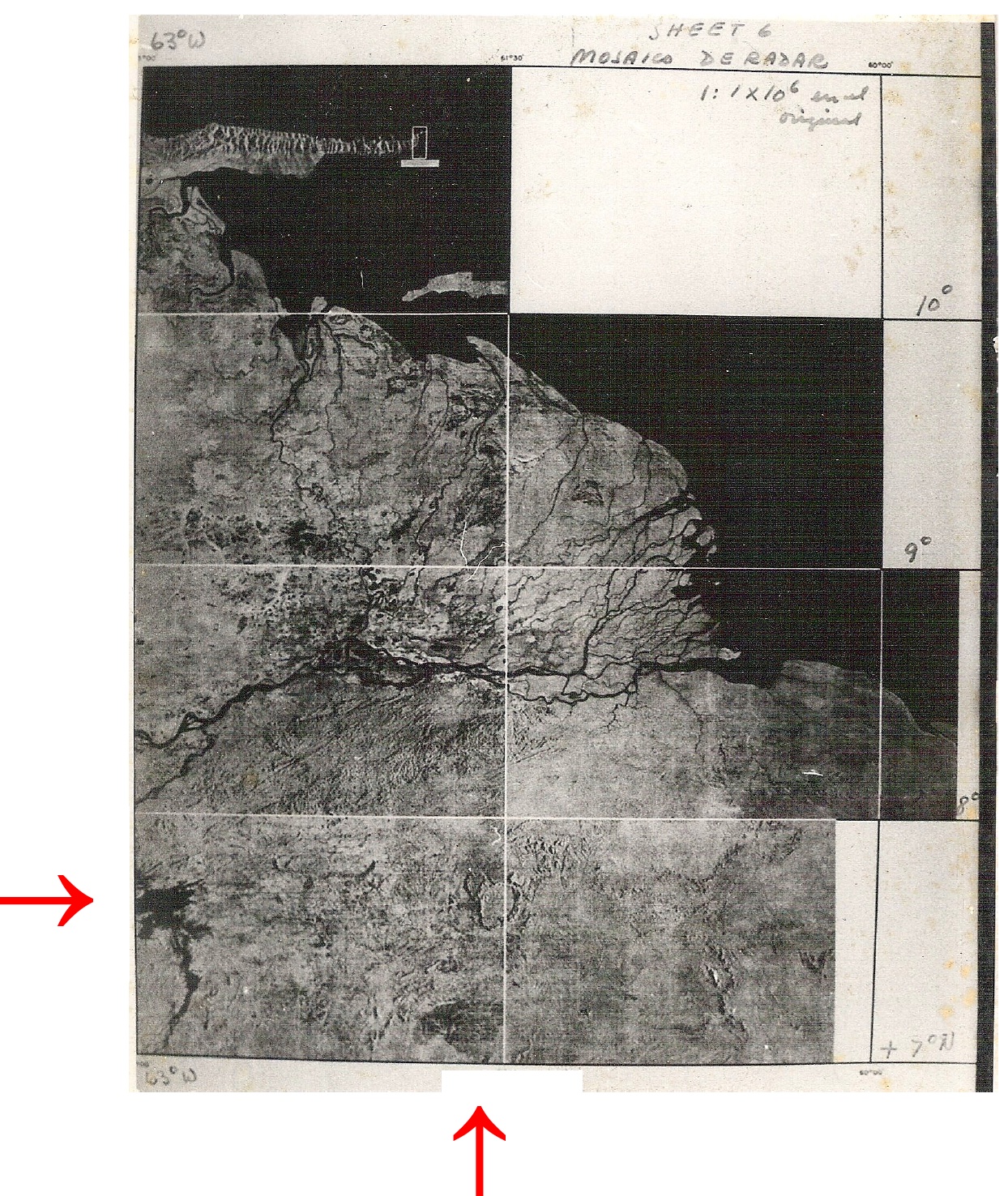}
    \caption{Historical airborne radar mosaic (1980s). The intersection of the red arrows marks the circular geomorphological feature investigated in this study. The image illustrates the appearance of the structure in radar backscatter data available prior to digital elevation models.}
    \label{fig:1}
\end{figure}

The material consists of:
\begin{itemize}
    \item correspondence between researchers (1980s)
    \item annotated radar imagery indicating suspected circular features
    \item aerial observational notes identifying two anomalous landforms (Fig.~\ref{fig:1})
\end{itemize}

\subsection{Modern topographic datasets}
\begin{figure}[htbp]
    \centering
    \includegraphics[width=0.8\linewidth]{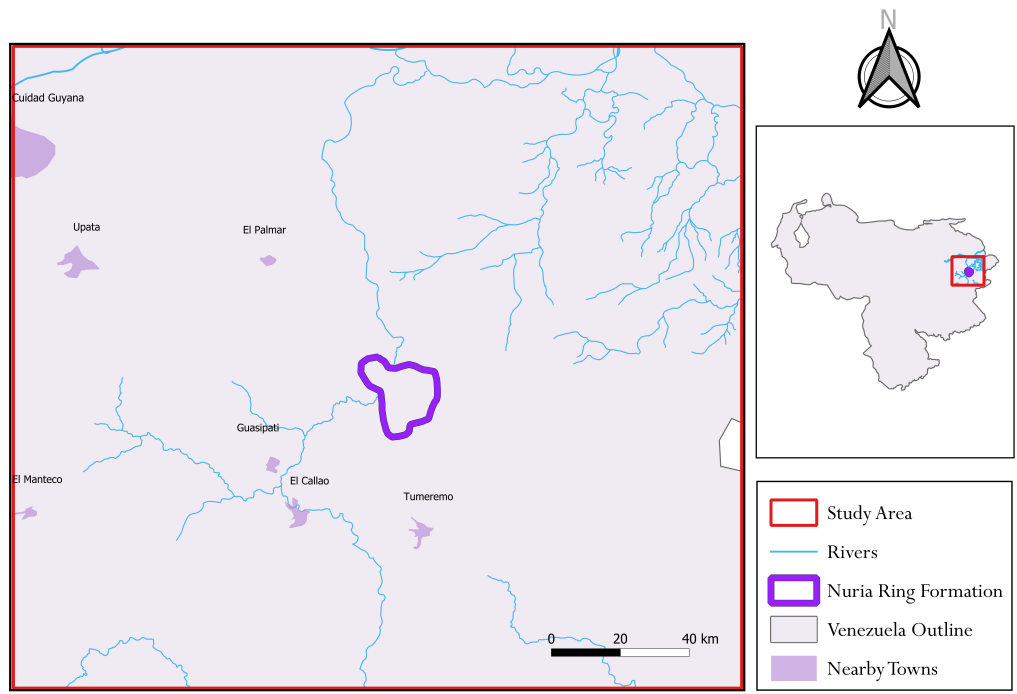}
    \caption{Geographic location of the study site within Venezuela. The map shows the position of the study area (red) in relation to the country's administrative boundaries.}
    \label{fig:2}
\end{figure}

The structure is located at approximately 7,6°N, 61,5°W within the northwestern Guiana Shield, Bolívar State, Venezuela. Its location within the national territory is shown in Fig.~\ref{fig:2}.


Topographic data were retrieved from the Shuttle Radar Topography Mission (SRTM) \citep{Farr2007SRTM}, which acquired radar elevation measurements during a Space Shuttle flight in February 2000. The mission employed both C-band ($\sim3''$, ~90\,m) and X-band ($\sim1''$, ~30\,m) interferometric radar.

\begin{landscape}
\begin{figure}[htbp]
    \centering
    \includegraphics[width=0.85\linewidth]{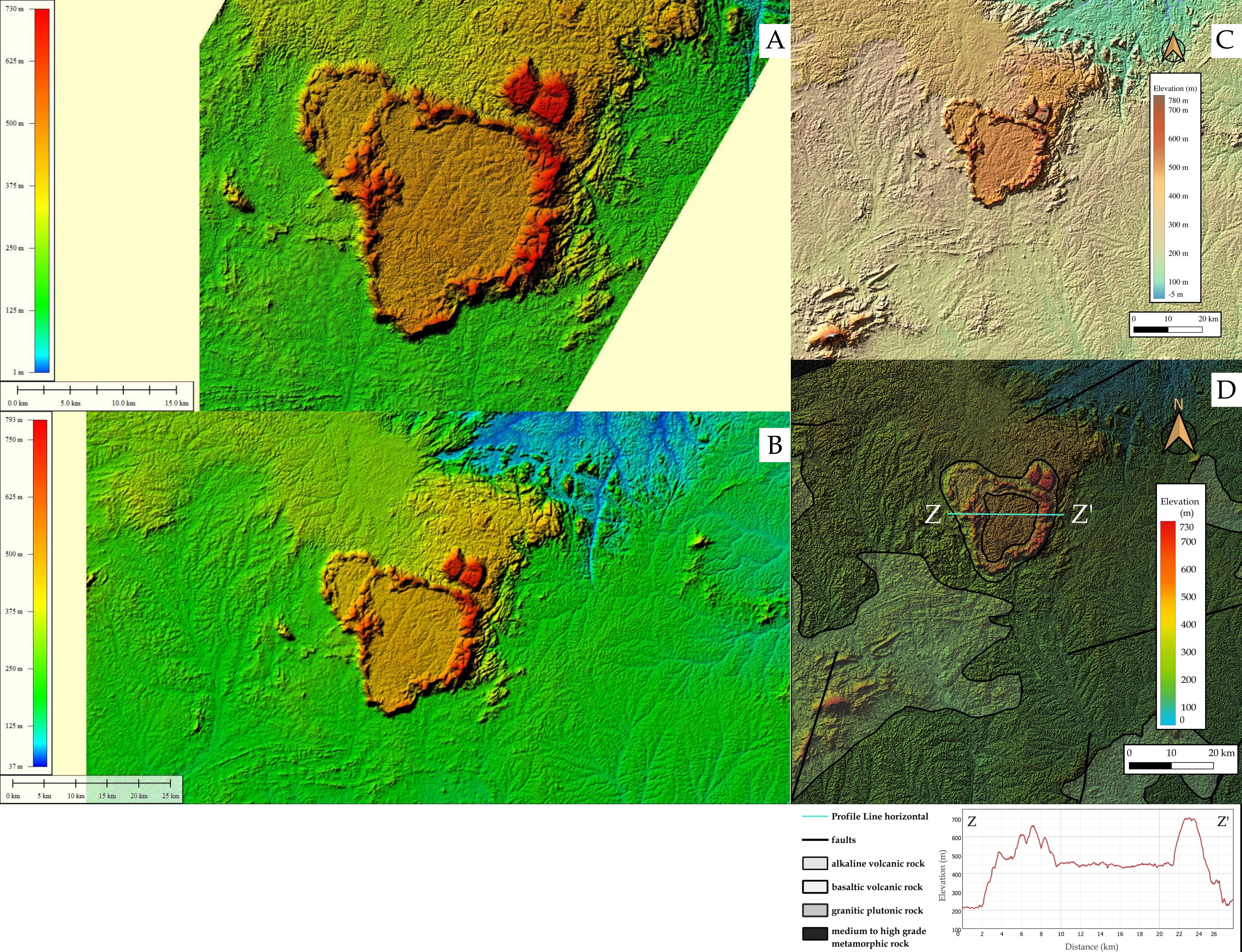}
    \caption{Digital elevation models of the study area derived from the Shuttle Radar Topography Mission (SRTM) and TanDEM-X. (A) SRTM X-band DEM (30\,m spatial resolution). (B) SRTM C-band DEM (90\,m spatial resolution). (C) TanDEM-X DEM (12\,m spatial resolution; only JPEG file available). (D) TanDEM-X Edited Digital Elevation Model (EDEM, 30\,m spatial resolution) with a transparent geological map overlay adapted from \cite{Gomez2019GeologicalMapSouthAmerica} and representative topographic profile (Z--Z$^\prime$). The profile shows the elevated annular morphology, a relatively flat interior, and $\sim$300–400\,m of relief. Elevation scale in meters above sea level.}
    \label{fig:3}
\end{figure}
\end{landscape}

SRTM digital elevation models (DEM) were obtained from public archives. The global C-band DEM was downloaded from the U.S. Geological Survey EarthExplorer portal\footnote{\url{https://earthexplorer.usgs.gov}}, while the higher-resolution X-band data covering the study area were retrieved from the German Aerospace Center (DLR) GeoService repository\footnote{\url{https://download.geoservice.dlr.de/SRTM_XSAR/}},\footnote{\url{https://geoservice.dlr.de/web/maps/srtm:x-sar}}. Both datasets are freely available and provide near-global topographic coverage.
The studied structure lies within X-band coverage; high-resolution topography was also available for morphological analysis (Fig.~\ref{fig:3}).
According to a recent evaluation of publicly available digital surface models, TerraSAR-X add-on for Digital Elevation Measurements (TanDEM-X) 12\,m data provide the highest spatial resolution \citep{Sefercik2025}. Therefore, Fig.~\ref{fig:3} includes this 12\,m file together with the TanDEM-X 30\,m dataset to illustrate the surface morphology in greater detail. The 12\,m and 30\,m resolution DEM were acquired through the Earth Observation on the Web Portal (EOWEB GeoPortal) and the EOC Geoservice of the Earth Observation Center (EOC) of the German Aerospace Center (DLR), respectively\footnote{\url{https://eoweb.dlr.de/egp/}},\footnote{\url{https://geoservice.dlr.de/web/maps/tdm:edem30}}. The 30\,m map was generated and hill-shaded for visualization using the Quantum Geographic Information System (QGIS) Software, version 3.40.2 \citep{QGIS} with azimuth = 315°, altitude = 45°, z-factor = 0.0005. No additional preprocessing, void filling, noise filtering, or datum harmonization procedures were applied beyond standard visualization and hillshading within QGIS.
Sensitivity tests were conducted by varying the illumination azimuth (280$^\circ$--400$^\circ$), illumination altitude angle (20$^\circ$--75$^\circ$), and z-factor (1 to $10^{-6}$). All parameter combinations yielded consistent large-scale geomorphological patterns. Furthermore, several perceptually uniform colormaps (Viridis, Cividis, Inferno, Magma, and Plasma) were evaluated in addition to grayscale hillshading. The principal geomorphological features remained unchanged across all visualization settings, demonstrating that the observed morphology is independent of illumination geometry, vertical exaggeration, and color representation. Although perceptually uniform colormaps are generally preferred for quantitative visualization, the rainbow-like colormap employed here enhances contrast between topographic units and facilitates recognition of the large-scale circular structure. We therefore conclude that the identified feature is not an artifact of the selected visualization parameters.

The historical radar mosaic primarily emphasized planform circularity because of illumination-dependent backscatter contrasts and the absence of quantitative topographic information. The SRTM C-band DEM resolved the broader annular plateau morphology, whereas the higher-resolution X-band and TanDEM-X datasets revealed discontinuous ridge geometry and the absence of a bowl-shaped depression or continuous raised rim. Consequently, progressively higher spatial resolution improved recognition of the structure as an intrusive geomorphological feature rather than an impact crater.

The authors do not suggest that the impact interpretation represented a formal or sustained scientific controversy, but rather an exploratory interpretation documented in surviving archival correspondence.


%



\section{Results}

\subsection{Morphology from SRTM data}

The structure has an approximate diameter of ~20–30 km measured across the annular plateau. Elevation differences between the surrounding terrain and the raised peripheral ring are approximately 300–400 m. The interior lacks a bowl-shaped depression and instead forms a relatively elevated plateau morphology (Fig.\,3(D)).

The SRTM elevation models reveal that the Venezuelan structure does not exhibit classical impact-crater morphology. 
Diagnostic impact-crater morphologies were not observed:
\begin{itemize}
    \item continuous raised rim
    \item central peak or peak ring
    \item symmetric ejecta distribution
    \item circular bowl-shaped depression
\end{itemize}

Impact crater morphology distinguishes between simple and complex craters \textcolor{black}{\citep{article}}. On Earth, simple craters generally occur at diameters below ~2–4\,km \citep{McCall2019} and are characterized by a well-defined circular planform and bowl-shaped geometry, although oblique impacts may produce elliptical crater geometries. In contrast, complex craters form above this threshold, exhibit larger diameters, reduced circularity, and commonly develop central uplifts or multi-ring structures \citep{Kenkmann2021}. The absence of these classical impact-crater morphologies, together with the regional geological context, is more consistent with a non-impact intrusive origin. 

\subsection{Regional geological context}
The structure lies between the Imataca Complex and the Pastora Series within the Guiana Shield, a Precambrian cratonic province. The topography aligns with regional tectonic fabrics including folds and faults. Such cratonic environments are known to host numerous intrusive ring complexes and weathered plateau structures that may mimic impact morphologies in planform geometry \citep{GibbsBarron1983GuianaShield}.

Subsequent literature review \citep{Wynn1991Nuria} identified the feature as the Nuria ring dike, an intrusive igneous body previously documented in geological surveys. The morphology observed in SRTM matches published geological maps.

Published geological descriptions \citep{Wynn1991Nuria} characterize the Nuria structure as an annular elevated plateau forming an amphitheater-like morphology, with a relatively flat interior and a raised peripheral ring approximately 300–400 m above the surrounding terrain and reaching elevations of 600–700 m above sea level. The central area consists of granitic gneiss belonging to the Supamo Complex, whereas the surrounding hills are composed of mafic intrusive rocks (diabase or gabbro). Lateritic bauxite deposits developed over these mafic lithologies have been reported, confirming prolonged weathering of intrusive material rather than excavation processes \citep{kalliokoski1965geology,Wynn1991Nuria,garcia1995geology,kroonenberg2019geology}.

Mafic intrusive rocks are particularly resistant to weathering \citep{Wörner2005}, which is why the circular structure protrudes. Such lithological relationships and alteration patterns are incompatible with impact-crater formation and instead indicate an intrusive ring-dike origin consistent with regional Guiana Shield geology. This is further supported by geological mapping shown in \citep{Kroonenberg2016}, where the Nuira formation is clearly depicted as mafic intrusive rock, present in diapiric tonalite-trondhjemite-granodiorite intrusive material. These lithological interpretations are derived from previously published regional geological studies rather than new field or laboratory analyses conducted within the present work.



\subsection{Comparative morphological analysis}

\begin{figure}[htbp]
    \centering
    \includegraphics[width=0.8\linewidth]{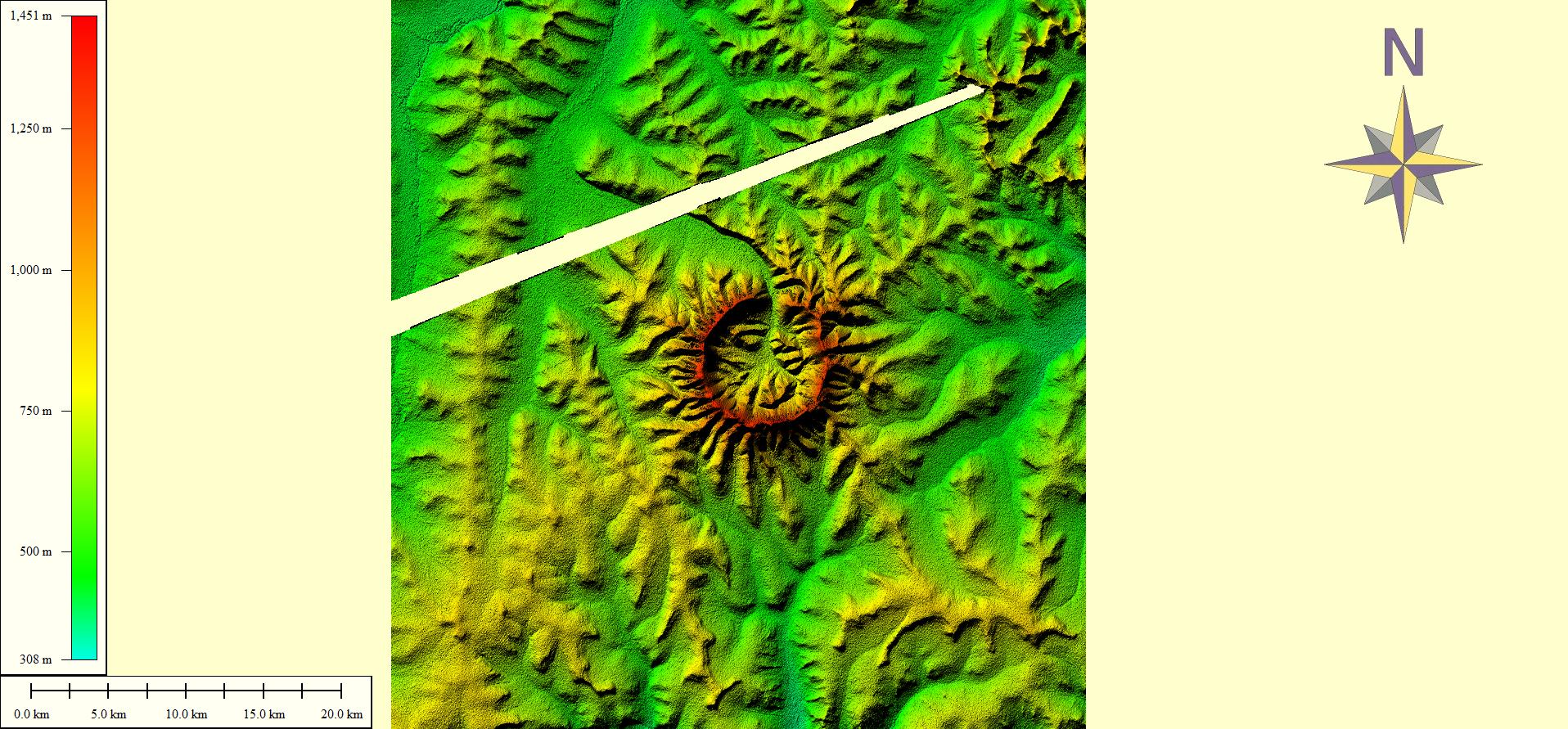}
    \caption{SRTM digital elevation model of the Kondyor massif (Russian Far East), a ring-shaped intrusive complex used as a morphological analogue. 90 m spatial resolution, elevation scale is in meters above mean sea level.}
    \label{fig:4}
\end{figure}

To illustrate the ambiguity of circular morphology, a comparison example of a known circular intrusion, the Kondyor massif in the Russian Far East located at approximately 57,35°N, 134,38°E, in Khabarovsk Krai, (Fig.~\ref{fig:4}) was used. Such structures may visually resemble impact craters in remote-sensing imagery but originate from intrusive magmatic processes rather than impact excavation.

\subsection{Agreement with modern geological mapping}

A regional geological map presented in a later study \citep{Reis2013Avanavero} outlines the Nuria structure with a geometry consistent with the morphology observed in the SRTM topography. This correspondence indicates that the feature belongs to the intrusive igneous complexes of the Guiana Shield. Comparable circular intrusions, such as the Kondyor massif \citep{Burg2009Kondyor}, demonstrate that ring-shaped landforms can originate from magmatic emplacement processes rather than impact excavation. Consequently, circular geometry alone cannot be considered diagnostic of impact origin.

The combined geomorphological observations presented here together with previously published lithological and geological mapping studies consistently support a non-impact intrusive origin for the structure.


\section{Discussion}

Similar misclassifications have historically occurred in early planetary mapping, where circular volcanic or tectonic structures were initially cataloged as impact craters prior to higher-resolution datasets. Classic examples exist across planetary exploration. Early lunar mapping classified several volcanic collapse structures and calderas as impact craters prior to orbital photography resolving their morphology \citep{Wilhelms1987Moon}. On Mars, circular volcanic edifices in the Tharsis region, such as Alba Patera and Uranius Patera, were initially difficult to distinguish from impact structures in early spacecraft imagery before later geomorphological analyses established their volcanic origin \citep{Carr2006Mars}. Radar observations from the Magellan mission later demonstrated that many circular features on Venus lacking ejecta blankets were volcanic coronae rather than impact craters \citep{Stofan1992Coronae}. 
This case illustrates how morphological interpretation without sufficient geological context may bias classification toward impact origin.
The present case represents a terrestrial analogue of the same interpretational bias produced by limited spatial resolution.
Therefore, integration of geological context and high-resolution topography provides a more robust basis for interpretation of circular structures on both terrestrial \citep{Gottwald2017} and planetary surfaces where such contextual information is available.

\subsection{Circular morphology and impact misidentification}
The historical interpretation relied primarily on planform geometry. However, circular appearance alone is insufficient to diagnose impact origin. Ancient terrestrial impact structures may lose several classical surface morphologies through prolonged erosion, tectonic deformation, weathering, \textcolor{black}{burial} or sedimentary infilling \textcolor{black}{\citep{Kenkmann2021}}.

An example of a well-known terrestrial case of planform geometry resulting from an impact event is the Nördlinger Ries structure in Bavaria, Germany. It is one of the best-preserved impact structures on Earth and shows definitive morphology and criteria \citep{Pohl1977, McCall2019, Wulf2020}. In contrast, some intrusive complexes form near-perfect circular rings despite non-impact origin. This can occur when magma intrudes and uplifts surrounding rock, later exposed by erosion. 

Such cases demonstrate a general principle important in planetary science: morphology must be supported by diagnostic criteria such as:

\begin{itemize}
    \item shock metamorphism (e.g. shatter cones, planar deformation features)
    \item brecciation (e.g. impact melt breccias)
    \item ejecta deposits (e.g. relics of projectile material)
    \item geophysical anomalies (e.g. electro- and magnetic, gravity, seismic faulting)
    \item {geochemical markers (e.g. high pressure mineral polymorphs)}
\end{itemize}

The Nuria formation is absent from catalogs of confirmed and possible terrestrial impact structures \citep{EarthImpactDatabase}, whose inclusion criteria require shock metamorphism or equivalent diagnostic evidence. Recrystallization of existing minerals or the formation of high pressure polymorphs are considered diagnostic geochemical markers \citep{Gottwald2022}; however, these are not observed in the geological record of the feature. It is important to note that criteria related to shock metamorphism and geochemical markers are definitive for confirming an impact origin, whereas other criteria are considered contributory \citep{EarthImpactDatabase}. This omission of these key criteria is consistent with the geological interpretation of the Nuria structure as a magmatic ring-dike complex rather than an impact crater. Many definitive diagnostic criteria, including shock metamorphism and high-pressure mineral polymorphs, cannot be evaluated using DEM data alone and instead require field investigation and laboratory analysis.

In the present case, none of these diagnostic criteria were identified by the SRTM morphology or by regional lithology and geological mapping. 

\subsection{Implications for planetary remote sensing}
The historical hypothesis anticipated a common modern challenge: identifying impact structures from limited-resolution datasets. Early radar imagery produced interpretations analogous to those encountered in orbital planetary mapping.

The resolution improvement from early radar imagery to SRTM data changed the interpretation qualitatively. This mirrors the evolution of planetary mapping from early spacecraft imagery to modern high-resolution orbital datasets. Unlike terrestrial environments, where erosion, vegetation cover, tectonic deformation, and weathering may significantly modify crater morphology over geological timescales, many planetary surfaces preserve impact-related landforms more directly because of reduced atmospheric and hydrological alteration.

\subsection{Value of historical scientific material}

The preserved correspondence documents a reasoning pathway typical of exploratory planetary geology:

\begin{itemize}
    \item geometric anomaly observed
    \item impact hypothesis proposed
    \item higher-resolution data required
    \item geological context resolves origin
\end{itemize}

Such documentation is uncommon in recorded material but provides insight into the development of remote-sensing interpretation methodology.

The study therefore functions as a terrestrial analogue demonstrating how improved spatial resolution can alter geomorphological classification in planetary datasets.

\subsection{Limitations}

The present study is based exclusively on remote sensing datasets and previously published geological literature. No field investigations, petrographic analyses, geochemical sampling, geophysical surveys, or laboratory mineralogical measurements were conducted. Consequently, the interpretation relies on geomorphological characteristics combined with regional geological context rather than direct subsurface verification.

The historical radar mosaic could not be quantitatively co-registered with modern DEM datasets because original acquisition metadata were unavailable. In addition, hillshade visualization may emphasize or suppress subtle landforms depending on illumination geometry, although the principal annular morphology remained consistent under alternative visualization conditions.

Therefore, the study should be interpreted primarily as a methodological and historiographical case study illustrating how circular geomorphological structures may be misidentified when interpretation relies predominantly on morphology and limited-resolution datasets.


\section{Conclusions}

Modern DEM datasets together with published geological studies consistently support interpretation of the Venezuelan structure as the Nuria ring dike rather than an impact crater. 

This study primarily documents a historical episode of geomorphological interpretation based on limited radar imagery prior to the availability of modern digital elevation data. The case highlights that circular morphology alone is not diagnostic of impact origin and illustrates how geological context combined with improved topographic visualization can clarify the origin of ambiguous circular landforms. 

More broadly, the study provides a terrestrial analogue for how increasing spatial resolution may alter geomorphological interpretation in planetary remote sensing and illustrates the value of integrating geomorphology, geological context, and topographic data when evaluating previously unresolved circular features.


\section*{Author Contributions}
MR conceived the study and wrote the draft. EK contributed to data analysis and to the writing of the manuscript. All authors reviewed and approved the final version of the manuscript.

\section*{Funding}
No specific funding

\section*{Acknowledgments}
The authors thank Manfred Gottwald for contributing with the SRTM data and geological interpretation and discussions. We also thank Gustavo Bruzual for preserving and providing the archival radar imagery—originally and privately circulated by a colleague, Lucrecia Maupomé (now deceased)—together with the accompanying correspondence that motivated this reanalysis. We also acknowledge the U.S. Geological Survey and the German Aerospace Center (DLR) for public provision of SRTM and TanDEM-X datasets. We thank the MPS IT center for its support installing QGIS. 

\bibliographystyle{Frontiers-Harvard} 
\bibliography{test}





\end{document}